\documentclass[journal, final, onecolumn, 12pt, letterpaper]{IEEEtran}
\usepackage[utf8]{inputenc}
\usepackage{amsmath}
\usepackage{amsfonts}
\usepackage{amssymb}
\usepackage{amsthm}
\usepackage{commath}
\usepackage{graphicx}
\usepackage{bm}
\usepackage{multirow}
\usepackage{epstopdf}
\usepackage{physics}
\usepackage{mathtools}
\usepackage{cases}
\usepackage{enumitem}
\usepackage{cite}
\usepackage{xcolor}
\usepackage{subfigure}

\usepackage{caption}
\allowdisplaybreaks
\newtheorem{theorem}{Theorem}
\newtheorem{definition}{Definition}
\newtheorem{lemma}[theorem]{Lemma}

\theoremstyle{definition}

\theoremstyle{definition}

\newcommand\scalemath[2]{\scalebox{#1}{\mbox{\ensuremath{\displaystyle #2}}}}

\makeatletter
\renewcommand*\env@matrix[1][\arraystretch]{%
  \edef\arraystretch{#1}%
  \hskip -\arraycolsep
  \let\@ifnextchar\new@ifnextchar
  \array{*\c@MaxMatrixCols c}}
\makeatother

\renewcommand{\arraystretch}{1.5}

\newcommand{\mli}[1]{\mathit{#1}}

\begin{document}

\title{Capacity of Generalized Discrete-Memoryless Push-to-Talk Two-Way Channels}

\author{Jian-Jia~Weng, Fady~Alajaji, and Tam{\'a}s~Linder%
\thanks{The authors are with the Department of Mathematics and Statistics, Queen’s University, Kingston, ON K7L 3N6, Canada (email: jian-jia.weng@queensu.ca, fa@queensu.ca, linder@mast.queensu.ca).}
\thanks{This work was supported in part by NSERC of Canada.}}

\maketitle
\begin{abstract}
In this report, we generalize Shannon's push-to-talk two-way channel (PTT-TWC) by allowing reliable full-duplex transmission as well as noisy reception in the half-duplex (PTT) mode. 
Viewing a PTT-TWC as two state-dependent one-way channels, we introduce a channel symmetry property pertaining to the one-way channels. 
Shannon's TWC capacity inner bound is shown to be tight for the generalized model under this symmetry property. 
We also analytically derive the capacity region, which is shown to be the convex hull of (at most) $4$ rate pairs. 
Examples that illustrate different shapes of the capacity region are given, and efficient transmission schemes are discussed via the examples.
\end{abstract}


\section{Introduction}
\begin{figure}[!t]
\centering
\includegraphics[draft=false, scale=0.7]{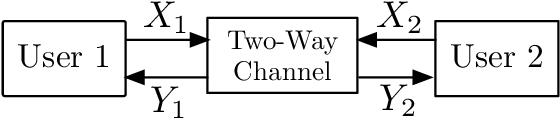}
\caption{Block diagram of a point-to-point TWC with channel inputs $X_1$ and $X_2$ and channel outputs $Y_1$ and $Y_2$.}
\label{fig:blkdiagram}
\end{figure}

Point-to-point two-way communication \cite{shannon1961} as depicted in Fig.~\ref{fig:blkdiagram} allows two users to simultaneously exchange information over a shared channel. 
Ideally, this enables cooperation between users to jointly improve the reliability of transmission via interactive adaptive coding. 
However, how each user can effectively maximize its individual transmission rate over the shared channel and concurrently provide sufficient feedback to help the other user's transmission is quite a challenging problem.
Although in the past two decades increased attention has been given to two-way channels (TWCs) \cite{meeuwissen1998, kramer1998, kramer2003, maor2006, gunduz2009, varshney2013, cheng2014, song2016, chaaban2017, jjw2017, jjw2018, palacio2018, sabag2018}, a single-letter characterization of the capacity region for general TWCs remains open. 
The aim of this report is to establish a capacity result for a generalized push-to-talk (PTT) TWC. 

Let $X_j\in\{0, 1, 2\}$ and $Y_j\in\{0, 1\}$ denote user-$j$'s channel input and output for $j=1, 2$, respectively.
Shannon's discrete-memoryless PTT-TWC (DM-PTT-TWC) \cite{shannon1961} as shown in Table~\ref{table:a} is a classic example where two-way simultaneous (i.e., full-duplex) transmission is completely unreliable and time-sharing between two one-way transmissions (i.e., half-duplex communication) is necessary to achieve capacity. 
As observed from the channel's marginal transition matrices in Tables~\ref{table:b} and~\ref{table:c}, user~1 can perfectly transmit a one-bit message to user~$2$ only when the channel input of user~$2$ is `0', and vice versa. 
Let $R_j$ denote the transmission rate of user~$j$ for $j=1, 2$. 
A simple time-sharing argument then gives the set of reliable transmission rate pairs $(R_1, R_2)=(\alpha, 1-\alpha)$, where $0\le\alpha\le 1$. 
Since there is no other way to transmit information reliably, that set of rate pairs clearly constitutes the boundary of the capacity region and thus determines capacity.\footnote{A formal proof of this statement via the Lagrange multiplier method can be found in \cite[Section~2.5.3]{meeuwissen1998}.} 

Inspired by Shannon's TWC setup, the PTT idea was extended to other multi-user channels such as PTT multiaccess channels \cite[Problem 14.7]{csiszar2011}, \cite{bierbaum1979}, switch-to-talk broadcast channels, and incompatible broadcast channels \cite[Section~V]{cover1972}.
In \cite{kramer2003}, a capacity result was established for a DM-PTT network with more than two users. 

In this report, we generalize Shannon's PTT-TWC by making two-way simultaneous transmission useful.
We also allow noisy reception in the half-duplex transmission and extend the channel input and output alphabets beyond ternary-input and binary-output.  
Viewing the PTT-TWC as two sets of one-way channels (one for each direction of transmission), we further introduce a channel symmetry property, which imposes on each transition matrix of the one-way channels a uniform structure, a weakly-symmetric structure\cite{cover2012}, and a capacity constraint. 
Under this symmetry property, we analytically derive the capacity region for the generalized PTT-TWCs. 
We also illustrate the possible different shapes of the capacity region and discuss efficient transmission strategies via examples.  

It is worth mentioning that a by-product of our derivation is a new way to show the tightness of Shannon's capacity inner bound \cite{shannon1961} which is complementary to prior methods in \cite{varshney2013, chaaban2017, jjw2018}. 
In fact, we find that none of these prior results imply that Shannon's inner bound is tight for Shannon's PTT-TWC (and for our general model under the symmetry property). 
We will discuss this issue later (in Section~III).   

The rest of this report is organized as follows. 
In Section~II, a brief review on the general TWC and the proposed DM-PTT-TWC models is given.
A capacity result for the proposed model is derived in Section~III. 
Examples are presented and qualitatively assessed in Section~IV, and conclusions are drawn in Section~V. 

\begin{table}[!t]
\caption{The full and marginal transition matrices of Shannon's PTT-TWC, where $X_j$ and $Y_j$ denote user-$j$'s channel input and output, respectively, $j=1, 2$. The rows and columns are indexed by the channel inputs and outputs, respectively.}
\centering
\begin{subtable}[$P_{Y_1, Y_2|X_1, X_2}${\cite[Table I]{shannon1961}}]{
\scalebox{1}{
\begin{tabular}{c||c|c|c|c}
$(X_1, X_2)$ & $(0, 0)$ & $(0, 1)$ & $(1, 0)$ & $(1, 1)$\\ 
\hline \hline 
$(0, 0)$ & $\frac{1}{4}$ & $\frac{1}{4}$ & $\frac{1}{4}$ & $\frac{1}{4}$ \\ 
\hline 
$(0, 1)$ & $\frac{1}{2}$ & $\frac{1}{2}$ & $0$ & $0$\\ 
\hline 
$(0, 2)$ & $0$ & $0$ & $\frac{1}{2}$ & $\frac{1}{2}$ \\ 
\hline \hline 
$(1, 0)$ & $\frac{1}{2}$ & $0$ & $\frac{1}{2}$ & $0$ \\ 
\hline 
$(1, 1)$ & $\frac{1}{4}$ & $\frac{1}{4}$ & $\frac{1}{4}$ & $\frac{1}{4}$ \\ 
\hline 
$(1, 2)$ & $\frac{1}{4}$ & $\frac{1}{4}$ & $\frac{1}{4}$ & $\frac{1}{4}$ \\ 
\hline \hline 
$(2, 0)$ & $0$ & $\frac{1}{2}$ & $0$ & $\frac{1}{2}$ \\ 
\hline 
$(2, 1)$ & $\frac{1}{4}$ & $\frac{1}{4}$ & $\frac{1}{4}$ & $\frac{1}{4}$ \\ 
\hline 
$(2, 2)$ & $\frac{1}{4}$ & $\frac{1}{4}$ & $\frac{1}{4}$ & $\frac{1}{4}$ \\ 
\end{tabular}}\label{table:a}}
\end{subtable}
\hfil
\begin{subtable}[$P_{Y_2|X_1, X_2}$]{
\scalebox{1}{
\begin{tabular}{c||c|c}
$(X_1, X_2)$ & $0$ & $1$ \\ 
\hline \hline 
$(0, 0)$ & $\frac{1}{2}$ & $\frac{1}{2}$ \\ 
\hline 
$(1, 0)$ & $1$ & $0$ \\ 
\hline 
$(2, 0)$ & $0$ & $1$ \\ 
\hline \hline 
$(0, 1)$ & $\frac{1}{2}$ & $\frac{1}{2}$ \\ 
\hline 
$(1, 1)$ & $\frac{1}{2}$ & $\frac{1}{2}$ \\ 
\hline 
$(2, 1)$ & $\frac{1}{2}$ & $\frac{1}{2}$ \\ 
\hline \hline 
$(0, 2)$ & $\frac{1}{2}$ & $\frac{1}{2}$ \\ 
\hline 
$(1, 2)$ & $\frac{1}{2}$ & $\frac{1}{2}$ \\ 
\hline 
$(2, 2)$ & $\frac{1}{2}$ & $\frac{1}{2}$ \\ 
\end{tabular}}\label{table:b}}
\end{subtable}
\hfil
\begin{subtable}[$P_{Y_1|X_1, X_2}$]{
\scalebox{1}{
\begin{tabular}{c||c|c}
$(X_1, X_2)$ & $0$ & $1$\\ 
\hline \hline 
$(0, 0)$ & $\frac{1}{2}$ & $\frac{1}{2}$\\ 
\hline 
$(0, 1)$ & $1$ & $0$\\ 
\hline 
$(0, 2)$ & $0$ & $1$\\ 
\hline \hline 
$(1, 0)$ & $\frac{1}{2}$ & $\frac{1}{2}$\\ 
\hline 
$(1, 1)$ & $\frac{1}{2}$ & $\frac{1}{2}$\\ 
\hline 
$(1, 2)$ & $\frac{1}{2}$ & $\frac{1}{2}$\\ 
\hline \hline 
$(2, 0)$ & $\frac{1}{2}$ & $\frac{1}{2}$\\ 
\hline 
$(2, 1)$ & $\frac{1}{2}$ & $\frac{1}{2}$\\ 
\hline 
$(2, 2)$ & $\frac{1}{2}$ & $\frac{1}{2}$\\ 
\end{tabular}}\label{table:c}}
\end{subtable}
\end{table}

\section{Preliminaries and Generalized\\ DM-PTT-TWC Model}
\subsection{General DM-TWC Model}
In point-to-point two-way communication, two users exchange messages $M_1$ and $M_2$ via $n$ channel uses. 
Messages $M_1$ and $M_2$ are assumed to be independent and uniformly distributed on the finite sets $\mathcal{M}_1 \triangleq \{ 1,2,..., 2^{\mli{nR}_1} \} $ and $\mathcal{M}_2 \triangleq \{ 1,2,...,2^{\mli{nR}_2} \}$, respectively, for some integers $\mli{nR}_1, \mli{nR}_2\ge 0$. 
Let $\mathcal{X}_j$ and $\mathcal{Y}_j$ be the channel input and output alphabets, respectively, for $j=1, 2$. 
For $i=1, 2, \dots, n$, let $X_{j,i} \in \mathcal{X}_j$ and $Y_{j,i}\in \mathcal{Y}_j$ denote the channel input and output of user $j$ at time $i$, respectively.  
Given the channel transition probability $P_{Y_1, Y_2|X_1, X_2}$, a TWC is said to be memoryless if $P_{Y_{1, i}, Y_{2, i}|X_1^i, X_2^i, Y_1^{i-1}, X_2^{i-1}}(y_{1, i}, y_{2, i}|x_1^i, x_2^i, y_1^{i-1}, y_2^{i-1})=P_{Y_1, Y_2|X_1, X_2}(y_{1, i}, y_{2, i}|x_{1, i}, x_{2, i})$ for all $i=1, 2, \dots, n$, where $x_j^i\triangleq (x_{j, 1}, x_{j, 2}, \dots, x_{j, i})$ and $y_j^{i-1}\triangleq (y_{j, 1}, y_{j, 2}, \dots, y_{j, i-1})$.  
A channel code for a DM-TWC is defined as follows. 

\begin{definition}{\label{def:CCforTWC}}
An $(n, R_1, R_2)$ code for a DM-TWC consists of two message sets $\mathcal{M}_1=\{1, 2, \dots, \allowbreak 2^{\mli{nR}_1}\}$ and $\mathcal{M}_2=\{1, 2, \dots, 2^{\mli{nR}_2}\}$, two sequences of encoding functions $f_1^n\triangleq (f_{1,1}, f_{1,2}, \dots, f_{1,n})$ and $f_2^n\triangleq (f_{2,1}, f_{2,2}, \dots, f_{2,n})$ such that $X_{1,1}= f_{1, 1}(M_1)$, $X_{2,1}=f_{2, 1}(M_2)$, $X_{1,i}=f_{1, i}(M_1, Y_1^{i-1})$, and $X_{2,i}=f_{2, i}(M_2, Y_2^{i-1})$
for $i=2, 3, \dots, n$, and two decoding functions $g_1$ and $g_2$ such that $\hat{M}_2=g_1(M_1, Y_1^n)$ and $\hat{M}_1=g_2(M_2, Y_2^n)$.
\end{definition}

When messages $M_1$ and $M_2$ are encoded via an $(n, R_1, R_2)$ channel code, the probability of decoding error is defined as $P^{(n)}_{\text{e}}(f_1^n, f_2^n, g_1, g_2)=\text{Pr}\{\hat{M}_1 \neq M_1\ \text{or}\ \hat{M}_2 \neq M_2\}.$

\begin{definition}
A rate pair $(R_1,R_2)$ is said to be achievable if there exists a sequence of $(n, R_1, R_2)$ codes such that $\lim_{n \to \infty} P^{(n)}_{\text{e}} =0$. 
The capacity region $\mathcal{C}$ is defined as the closure of all achievable rate pairs. 
\end{definition}

To date, a computable single-letter expression for the capacity region of general DM-TWCs has not been found. 
Capacity bounds such as \cite{shannon1961, han1984, zhang1986, hekstra1989} still play crucial roles in studying transmission problems over DM-TWCs.  
Let $\mathcal{R}(P_{X_1,X_2},P_{Y_1,Y_2|X_1,X_2})$ denote the set of rate pairs $(R_1,R_2)$ with $R_1 \le I(X_1;Y_2|X_2)$ and $R_2 \le I(X_2;Y_1|X_1)$, 
 where the joint distribution of all random variables is given by $P_{X_1, X_2, Y_1, Y_2}=P_{X_1,X_2}\cdot P_{Y_1,Y_2|X_1,X_2}$. 
Shannon in \cite{shannon1961} showed that the capacity region of a DM-TWC with transition probability $P_{Y_1,Y_2|X_1,X_2}$ is inner bounded by 
 \begin{align*}
 \scalemath{1}{\mathcal{C}_\text{I} (P_{Y_1,Y_2|X_1,X_2}) \triangleq \scalemath{1}{\overline{\text{co}} \left( \bigcup_{P_{X_1}P_{X_2}} \mathcal{R}(P_{X_1}P_{X_2}, P_{Y_1,Y_2|X_1,X_2}) \right)}},
 \end{align*}
 and outer bounded by
 \begin{align*}
\scalemath{1}{\mathcal{C}_\text{O} (P_{Y_1,Y_2|X_1,X_2}) \triangleq \scalemath{1}{\bigcup_{P_{X_1,X_2}} \mathcal{R}(P_{X_1,X_2}, P_{Y_1,Y_2|X_1,X_2})}},
  \end{align*}
where $\overline{\text{co}}$ denotes taking the closure of the convex hull. 
An alternative expression of $\mathcal{C}_\text{I}$ without the convex hull operation can be obtained by introducing an auxiliary random variable \cite[Proposition~17.2]{kim2011}.  

In general, $\mathcal{C}_\text{I}$ and $\mathcal{C}_\text{O}$ do not coincide, but various sufficient conditions that imply the tightness of $\mathcal{C}_\text{I}$ have been proposed in \cite{shannon1961, chaaban2017, jjw2018}. 
However, these conditions only apply to DM-TWCs for which the convex hull operation is unnecessary in obtaining $\mathcal{C}_\text{I}$, thus failing to determine the capacity region for channels requiring this operation, such as Shannon's PTT-TWC. 
In this report, we address this issue for a generalized DM-PTT-TWC model.   

\subsection{Generalized DM-PTT-TWC Model}
For $j=1, 2$, let $\mathcal{X}_j\triangleq\{0, 1, \dots, r_j-1\}$ and $\mathcal{Y}_j\triangleq\{0, 1, \dots, s_j-1\}$, where $r_j\ge 3$ and $s_j\ge 2$ (to avoid trivial cases). 
Without loss of generality, we set $X_1=0$ and $X_2=0$ as the signals for the ``PTT mode''. 
For $j=1, 2$, let $\bm{v}_j$ denote the length-$s_j$ row vector with all entries equal to $1/s_j$.
Also, let $\bm{Q}_{j, x_k}$ denote a $(r_j-1)\times s_k$ channel transition matrix with capacity $C_{j, x_k}$ for $j, k=1, 2$ with $j\neq k$ and $x_k\in\mathcal{X}_k$. 
An $(r_1, r_2, s_1, s_2)$ generalized DM-PTT-TWC with transition probability $P_{Y_1, Y_2|X_1, X_2}$ is defined by imposing the following structure for the marginal channel transition matrices $[P_{Y_j|X_1, X_2}(\cdot|\cdot, \cdot)]$ (where the rows and columns are indexed by the channel inputs and outputs, respectively): for all $x_2\in\mathcal{X}_2$, 
\begin{IEEEeqnarray}{l}
[P_{Y_2|X_1, X_2}(\cdot|\cdot, x_2)]=
\Bigl(\begin{smallmatrix}\bm{v}_2\\
\bm{Q}_{1, x_2}\\
\end{smallmatrix}\Bigr),\nonumber
\end{IEEEeqnarray}
and for all $x_1\in\mathcal{X}_1$,
\begin{IEEEeqnarray}{l}
[P_{Y_1|X_1, X_2}(\cdot|x_1, \cdot)]=\Bigl(\begin{smallmatrix} 
\bm{v}_1\\
\bm{Q}_{2, x_1}\\
\end{smallmatrix}\Bigr).\nonumber
\end{IEEEeqnarray}
We remark that the above structures do not imply the property $P_{Y_1, Y_2|X_1, X_2}=P_{Y_1|X_1, X_2}\cdot P_{Y_2|X_1, X_2}$.  

Unlike Shannon's original PTT-TWC, our proposed model considers both perfect and noisy reception in the PTT mode and allows reliable full-duplex transmission.
Shannon's PTT-TWC can be recovered by setting $(r_1, r_2, s_1, s_2)=(3, 3, 2, 2)$, $\bm{Q}_{j, 0}=\bm{I}_2$, and $\bm{Q}_{j, 1}=\bm{Q}_{j, 2}=\frac{1}{2}\cdot\bm{1}_{2\times 2}$ for $j=1, 2$, where $\bm{I}_2$ and $\bm{1}_{2\times 2}$ denote the $2\times 2$ identity and all-one matrices, respectively, and the overall channel transition probability can be obtained as $P_{Y_1, Y_2|X_1, X_2}=P_{Y_1|X_1, X_2}\cdot P_{Y_2|X_1, X_2}$. 

\section{Capacity Region of Generalized DM-PTT-TWCs with a Symmetry Property}
The capacity region of an $(r_1, r_2, s_1, s_2)$ DM-PTT-TWC is generally unknown. 
Below, we show that the capacity region can be analytically determined when the marginal channels exhibit the following symmetry property:\smallskip\\
\noindent \textbf{Channel Symmetry Property for Generalized PTT-TWCs}: for $j, k=1, 2$ with $j\neq k$, $\bm{Q}_{j, x_k}$'s are weakly-symmetric\footnote{A channel is said to be weakly-symmetric if its transition matrix has identical column sums and its rows are permutations of each other \cite[Section 7.2]{cover2012}; for  such a channel, the mutual information is maximized by the uniform input distribution. We note that for more general symmetric transition matrices for which mutual information is maximized by the uniform input distribution (e.g. quasi-symmetric channels \cite{alajaji2018}), Theorem~1 does not necessarily hold.} for all $x_k\in\mathcal{X}_k$ and $C_{j, x_k}=C_{j, 1}$ for all $x_k\neq 0$.\smallskip

Letting $\mathbf{1}\{\cdot\}$ denote indicator function, and letting $P^{\text{U}_0}_{\mathcal{X}_j}$ denote the probability distribution that assigns zero probability mass to $X_j=0$ and is uniform over the set $\mathcal{X}_j\backslash \{0\}$, $j=1, 2$, we define six rate pairs and their associated input distributions for the generalized PTT-TWC with the above symmetry property as follows:
\begin{itemize}
\item $\bm{R}^*_1\triangleq (0, 0)$, $P_{X_1, X_2}(x_1, x_2)=\mathbf{1}\{x_1=0\}\cdot\mathbf{1}\{x_2=0\}$;
\item $\bm{R}^*_2\triangleq(C_{1, 1}, C_{2, 1})$, $P_{X_1, X_2}=P^{\text{U}_0}_{\mathcal{X}_1}\cdot P^{\text{U}_0}_{\mathcal{X}_2}$;
\item $\bm{R}^*_3\triangleq(C_{1, 0}, 0)$, $P_{X_1, X_2}(x_1, x_2)=P^{\text{U}_0}_{\mathcal{X}_1}(x_1)\cdot\mathbf{1}\{x_2=0\}$;
\item $\bm{R}^*_4\triangleq(0, C_{2, 0})$, $P_{X_1, X_2}(x_1, x_2)=\mathbf{1}\{x_1=0\}\cdot P^{\text{U}_0}_{\mathcal{X}_2}(x_2)$;
\item $\bm{R}^*_5\triangleq(C_{1, 1}, 0)$, $P_{X_1, X_2}(x_1, x_2)=P^{\text{U}_0}_{\mathcal{X}_1}(x_1)\cdot\mathbf{1}\{x_2=1\}$;
\item $\bm{R}^*_6\triangleq(0, C_{2, 1})$, $P_{X_1, X_2}(x_1, x_2)=\mathbf{1}\{x_1=1\}\cdot P^{\text{U}_0}_{\mathcal{X}_2}(x_2)$.
\end{itemize}
Note that the $\bm{R}^*_l$'s are all attained via independent inputs.  

\begin{theorem}
For an $(r_1, r_2, s_1, s_2)$ DM-PTT-TWC that satisfies the above channel symmetry property, Shannon's inner bound is tight and the capacity region can be determined by taking the convex hull of $\bm{R}^*_1$, $\bm{R}^*_2$, $\max(\bm{R}^*_3, \bm{R}^*_5)$, and $\max(\bm{R}^*_4, \bm{R}^*_6)$.\footnote{We set $\max(\bm{A}, \bm{B})=\bm{B}$ iff $\bm{A}$ is upper-bounded component-wise by $\bm{B}$.} 
\end{theorem}

The idea behind the proof of Theorem~1 is to show that any rate pair in Shannon's outer bound region $\mathcal{C}_\text{O}$ can be upper-bounded component-wise by another rate pair that is a convex combination of the $\bm{R}^*_l$'s. 
More specifically, depending on the value of $C_{j, x_k}$'s, we can use the four rate pairs: $\bm{R}^*_1$, $\bm{R}^*_2$, $\max(\bm{R}^*_3, \bm{R}^*_5)$, and $\max(\bm{R}^*_4, \bm{R}^*_6)$, to upper-bound any rate pair in $\mathcal{C}_\text{O}$ and hence determine the capacity region. 
Here, we only prove the case where $\bm{R}^*_3=\max(\bm{R}^*_3, \bm{R}^*_5)$ and $\bm{R}^*_4=\max(\bm{R}^*_4, \bm{R}^*_6)$.   
The same argument can be used to prove other cases, and hence the details are omitted. 

\begin{IEEEproof}[Proof of Theorem~1]
Given any $P_{X_1, X_2}$, we bound the associated rate pair $(I(X_1; Y_2|X_2), I(X_2; Y_1|X_1))$ as follows:
\begin{IEEEeqnarray}{rCl}
I(X_1; Y_2|X_2) &=& \sum_{x_2=0}^{r_2-1}P_{X_2}(x_2)\cdot I(X_1; Y_2|X_2=x_2)\label{eq5}\\
&\le &  \sum_{x_2=0}^{r_2-1}P_{X_2}(x_2)\cdot \left[(1-P_{X_1|X_2}(0|x_2))\cdot C_{1, x_2}\right]\label{eq1}\\
& =&  (P_{X_2}(0)-P_{X_1, X_2}(0, 0))\cdot C_{1, 0} + \sum_{x_2\neq 0}(P_{X_2}(x_2)-P_{X_1, X_2}(0, x_2))\cdot C_{1, x_2}\nonumber\\
&=&  (P_{X_2}(0)-P_{X_1, X_2}(0, 0))\cdot C_{1, 0} + \sum_{x_2\neq 0}(P_{X_2}(x_2)-P_{X_1, X_2}(0, x_2))\cdot C_{1, 1}\nonumber\\
& &\ \ \quad + \underbrace{(P_{X_1}(0)-P_{X_1, X_2}(0, 0))\cdot 0 + P_{X_1, X_2}(0)\cdot 0}_{=0}\label{eq2},\IEEEeqnarraynumspace
\end{IEEEeqnarray}
where \eqref{eq1} follows from Lemma~\ref{lma:rateloss} in the Appendix and \eqref{eq2} holds since $C_{1, x_2}=C_{1, 1}$ for all $x_2\neq 0$. 
Similarly, we have
\begin{IEEEeqnarray}{rCl}
I(X_2; Y_1|X_1)&=& \sum_{x_1=0}^{r_1-1}P_{X_1}(x_1)\cdot I(X_2; Y_1|X_1=x_1)\label{eq6}\\
 &\le & \sum_{x_1=0}^{r_1-1}P_{X_1}(x_1)\cdot \left[(1-P_{X_2|X_1}(0|x_1))\cdot C_{2, x_1}\right]\nonumber\\
& = & (P_{X_1}(0)-P_{X_1, X_2}(0, 0))\cdot C_{2, 0}  + \sum_{x_1\neq 0}(P_{X_1}(x_1)-P_{X_1, X_2}(x_1, 0))\cdot C_{2, x_1}\nonumber\\
&=&  (P_{X_1}(0)-P_{X_1, X_2}(0, 0))\cdot C_{2, 0} + \sum_{x_1\neq 0}(P_{X_1}(x_1)-P_{X_1, X_2}(x_1, 0))\cdot C_{2, 1}\nonumber\\
& & \ \ \quad + \underbrace{(P_{X_2}(0)-P_{X_1, X_2}(0, 0))\cdot 0 + P_{X_1, X_2}(0)\cdot 0}_{=0}.\IEEEeqnarraynumspace\label{eq3}
\end{IEEEeqnarray}
Note that \eqref{eq2} and \eqref{eq3} and the fact that $\sum_{x_2\neq 0}(P_{X_2}(x_2)-P_{X_1, X_2}(0, x_2))=\sum_{x_1\neq 0}(P_{X_1}(x_1)-P_{X_1, X_2}(x_1, 0))$ imply that the pair $(I(X_1; Y_2|X_2), I(X_2; Y_1|X_1))$ is upper-bounded component-wise by 
\begin{IEEEeqnarray}{l}
P_{X_1, X_2}(0){\cdot}\bm{R}^*_1 + \Bigg[\sum_{x_1\neq 0}P_{X_1}(x_1){-}P_{X_1, X_2}(x_1, 0)\Bigg]{\cdot}\bm{R}^*_2+\nonumber\\
\ \ \ \ \left[P_{X_2}(0){-}P_{X_1, X_2}(0, 0)\right]{\cdot}\bm{R}^*_3+\left[P_{X_1}(0){-}P_{X_1, X_2}(0, 0)\right]{\cdot}\bm{R}^*_4\label{eq4}.\nonumber
\end{IEEEeqnarray}
Since the coefficients of the above four rate pairs sum to one, any rate pair in $\mathcal{C}_\text{O}$ is outer bounded by some convex combination of $\bm{R}^*_1$, $\bm{R}^*_2$, $\bm{R}^*_3$, and $\bm{R}^*_4$. 
Since the four rate pairs are achievable via independent inputs, we conclude that Shannon's inner bound is tight. 
\end{IEEEproof}

Clearly, the capacity region of Shannon's PTT-TWC can be easily determined via Theorem~1 without using the time-sharing argument \cite{shannon1961} or the Lagrange multiplier method \cite{meeuwissen1998}.

Moreover, we note that \eqref{eq5} can be interpreted as the average amount of information sent over a set of state-dependent one-way channels $\{[P_{Y_2|X_1, X_2}(\cdot|\cdot, x_2)]{:}\ x_2\in\mathcal{X}_2\}$, where the channel input, output, and state, correspond to $X_1$, $Y_2$, and $X_2$, respectively.
Thus, user-2's input distribution $P_{X_2}$ not only carries its own message but also determines how often each one-way channel can be used for user~1. 
The same interpretation also applies to \eqref{eq6}.  
Clearly, the best channel input distribution for one user may not create the most favorable one-way channel allocation for the other user, necessitating a rate trade-off between the two users' transmissions.  

Quantifying the trade-off is often the most involved part of determining the capacity region of general TWCs. 
The prior approach to tackle the problem is to exploit (when they exist) channel symmetry or invariance properties so that for any $P_{X_1, X_2}=P_{X_2}\cdot P_{X_1|X_2}$, one can always find a $\tilde{P}_{X_1}$ such that $\mathcal{R}(P_{X_1, X_2}, P_{Y_1, Y_2|X_1, X_2})\subseteq \mathcal{R}(\tilde{P}_{X_1}\cdot P_{X_2}, P_{Y_1, Y_2|X_1, X_2})$  \cite{shannon1961, chaaban2017, jjw2018}. 
However, this approach fails here since such $\tilde{P}_{X_1}$ may not exist for each $P_{X_1, X_2}$. 
This observation can be illustrated via Shannon's PTT-TWC as one can see that no single independent input distribution can achieve the rate pair $(R_1, R_2)=(\alpha, 1-\alpha)$, where $0<\alpha<1$. 
It is thus of interest to exploit other symmetry property as the one presented at the beginning of the section that allows us to show $\mathcal{C}_\text{O}\subseteq \mathcal{C}_\text{I}$ directly. 

\section{Examples and Discussion}
In the last section, we proved the tightness of Shannon's inner bound for a class of generalized DM-PTT-TWCs. 
The capacity result in Theorem~1 suggests a way to use different state-dependent one-way channels to optimize bi-directional transmission rates. 
In what follows, we illustrate all possible shapes of the capacity region via examples and discuss the optimal transmission strategy behind each result.   

Let $(r_1, r_2, s_1, s_2)=(3, 3, 3, 3)$. Consider the generalized PTT-TWC with the parameterized marginal transition matrices as shown in Table~II and the following settings:\medskip

\noindent\underline{Setting 1}: $(a, b, c, d)=(0, 0.15, 0, 0.15)$ $\Rightarrow$ 
\[C_{j, 0}=0.6667>C_{j, x_k}=0.1539\] 
for $j, k=1, 2$ with $j\neq k$ and all $x_k\neq 0$;\smallskip\\
\noindent\underline{Setting 2}: $(a, b, c, d)=(0, 0.05, 0, 0.01)$ $\Rightarrow$ 
\begin{IEEEeqnarray}{rCl}
C_{1, 0}=0.6667 &>& C_{1, x_2}=0.4105\nonumber\\
C_{2, 0}=0.6667 &>& C_{2, x_1}=0.5918\nonumber
\end{IEEEeqnarray}
for all $x_1\neq 0$ and $x_2\neq 0$;\smallskip\\ 
\noindent\underline{Setting 3}: $(a, b, c, d)=(0.1, 0, 0, 0.01)$ $\Rightarrow$ 
\begin{IEEEeqnarray}{rCl}
C_{1, 0}=0.2601 &<& C_{1, x_2}=0.6667\nonumber\\
C_{2, 0}=0.6667 &>& C_{2, x_1}=0.5918\nonumber
\end{IEEEeqnarray}
for all $x_1\neq 0$ and $x_2\neq 0$;\smallskip\\
\noindent\underline{Setting 4}: $(a, b, c, d)=(0.1, 0, 0.2, 0.05)$ $\Rightarrow$ 
\begin{IEEEeqnarray}{rCl}
C_{1, 0}=0.2601 &<& C_{1, x_2}=0.6667\nonumber\\
C_{2, 0}=0.0791 &<& C_{2, x_1}=0.4105\nonumber
\end{IEEEeqnarray}
for all $x_1\neq 0$ and $x_2\neq 0$.\medskip\\
Note that, unlike for Shannon's original PTT-TWC, reliable full-duplex transmission is possible in the above settings since $C_{j, x_k}>0$ for all $j, k=1, 2$ with $j\neq k$ and $x_k\in\mathcal{X}_k$. 

\begin{table}[!t]
\caption{Marginal transition matrices of a generalized PTT-TWC, where $0\le a, b, c, d\le \frac{2}{3}$.}
\centering
\begin{subtable}[$P_{Y_2|X_1, X_2}$]{
\scalebox{1}{
\begin{tabular}{c||c|c|c}
$(X_1, X_2)$ & $0$ & $1$ & $2$ \\ 
\hline \hline 
$(0, 0)$ & $\frac{1}{3}$ & $\frac{1}{3}$ & $\frac{1}{3}$ \\ 
\hline 
$(1, 0)$ & $\frac{2}{3}-a$ & $a$ & $\frac{1}{3}$ \\ 
\hline 
$(2, 0)$ & $a$ & $\frac{2}{3}-a$ & $\frac{1}{3}$ \\ 
\hline \hline 
$(0, 1)$ & $\frac{1}{3}$ & $\frac{1}{3}$ & $\frac{1}{3}$ \\ 
\hline 
$(1, 1)$ & $\frac{2}{3}-b$ & $b$ & $\frac{1}{3}$ \\ 
\hline 
$(2, 1)$ & $b$ & $\frac{2}{3}-b$ & $\frac{1}{3}$ \\ 
\hline \hline 
$(0, 2)$ & $\frac{1}{3}$ & $\frac{1}{3}$ & $\frac{1}{3}$ \\ 
\hline 
$(1, 2)$ & $\frac{2}{3}-b$ & $b$ & $\frac{1}{3}$ \\ 
\hline 
$(2, 2)$ & $b$ & $\frac{2}{3}-b$ & $\frac{1}{3}$ \\ 
\end{tabular}}}
\end{subtable}
\hfil
\begin{subtable}[$P_{Y_1|X_1, X_2}$]{
\scalebox{1}{
\begin{tabular}{c||c|c|c}
$(X_1, X_2)$ & $0$ & $1$ & $2$ \\ 
\hline \hline 
$(0, 0)$ & $\frac{1}{3}$ & $\frac{1}{3}$ & $\frac{1}{3}$ \\ 
\hline 
$(0, 1)$ & $\frac{2}{3}-c$ & $c$ & $\frac{1}{3}$ \\ 
\hline 
$(0, 2)$ & $c$ & $\frac{2}{3}-c$ & $\frac{1}{3}$ \\ 
\hline \hline 
$(1, 0)$ & $\frac{1}{3}$ & $\frac{1}{3}$ & $\frac{1}{3}$ \\ 
\hline 
$(1, 1)$ & $\frac{2}{3}-d$ & $d$ & $\frac{1}{3}$ \\ 
\hline 
$(1, 2)$ & $d$ & $\frac{2}{3}-d$ & $\frac{1}{3}$ \\ 
\hline \hline 
$(2, 0)$ & $\frac{1}{3}$ & $\frac{1}{3}$ & $\frac{1}{3}$ \\ 
\hline 
$(2, 1)$ & $\frac{2}{3}-d$ & $d$ & $\frac{1}{3}$ \\ 
\hline 
$(2, 2)$ & $d$ & $\frac{2}{3}-d$ & $\frac{1}{3}$ \\ 
\end{tabular}}}
\end{subtable}
\end{table}

\begin{figure*}
\centerline{\subfigure[Setting 1 (unused rate pairs: $\bm{R}_2=(0.1539, 0.1539)$, $\bm{R}_5=(0.1539, 0)$, and $\bm{R}_6=(0, 0.1539)$)]{\includegraphics[draft=false, width=3in]{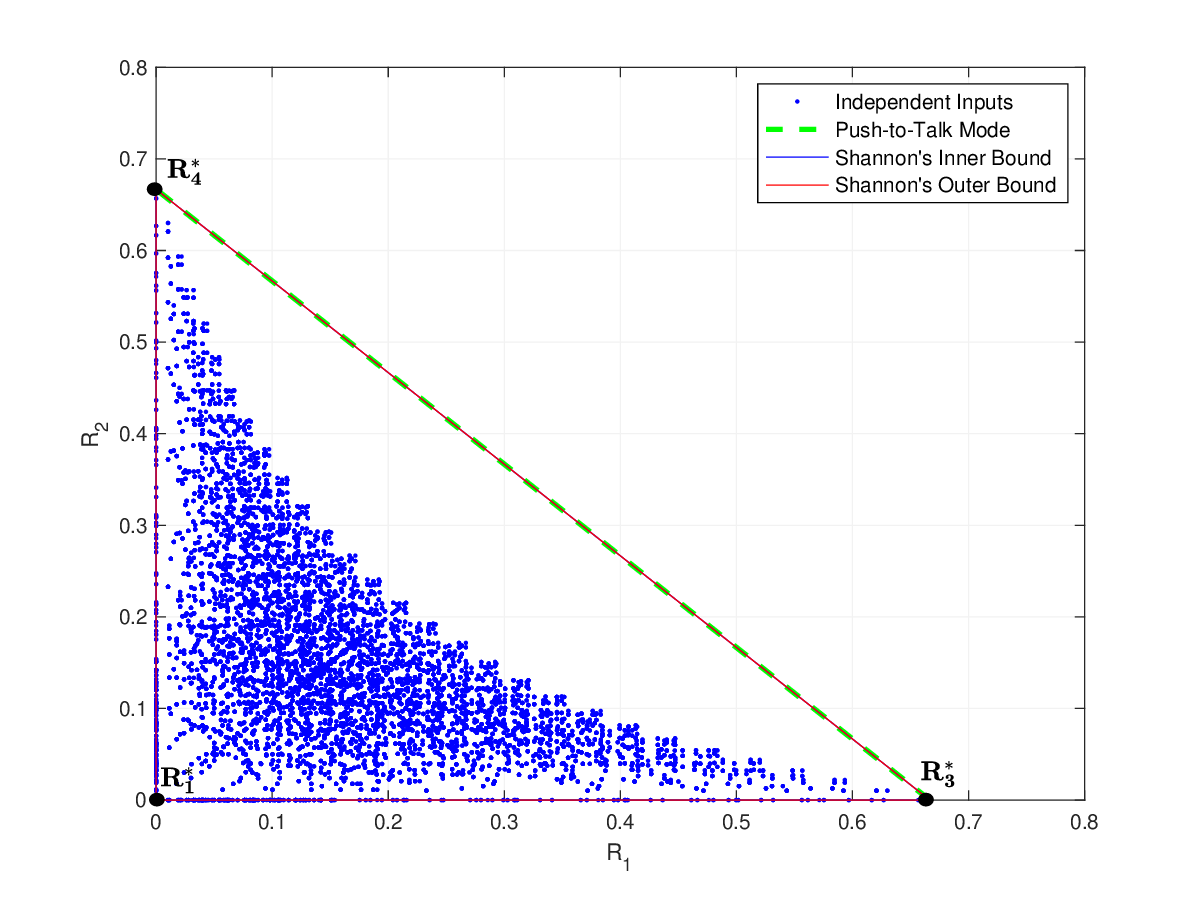}
\label{fig:1}}
\hfil
\subfigure[Setting 2 (unused rate pairs: $\bm{R}_5=(0.4105, 0)$ and $\bm{R}_6=(0, 0.5918)$)]{\includegraphics[draft=false, width=3in]{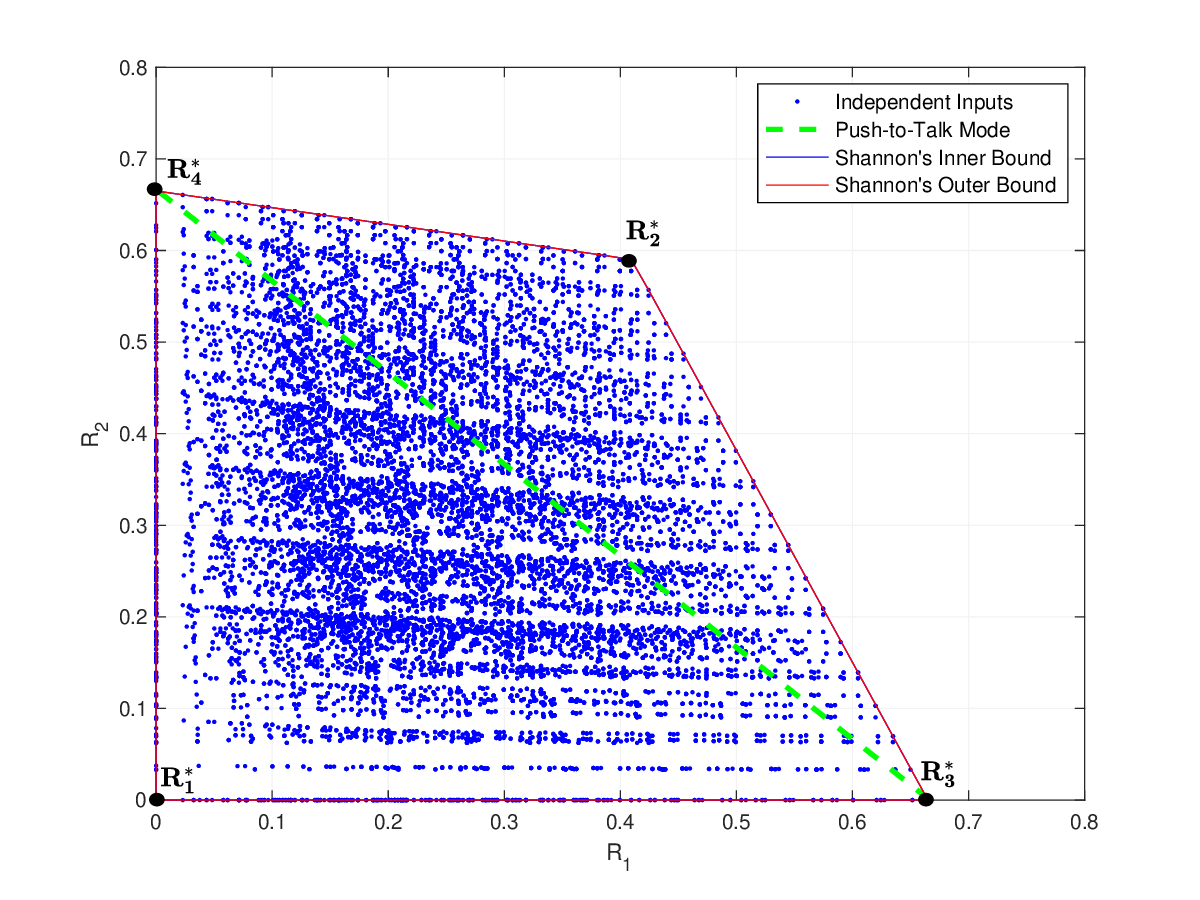}
\label{fig:2}}}\par
\centerline{
\subfigure[Setting 3 (unused rate pairs: $\bm{R}_3=(0.2601, 0)$ and $\bm{R}_6=(0, 0.5918)$)]{\includegraphics[draft=false, width=3in]{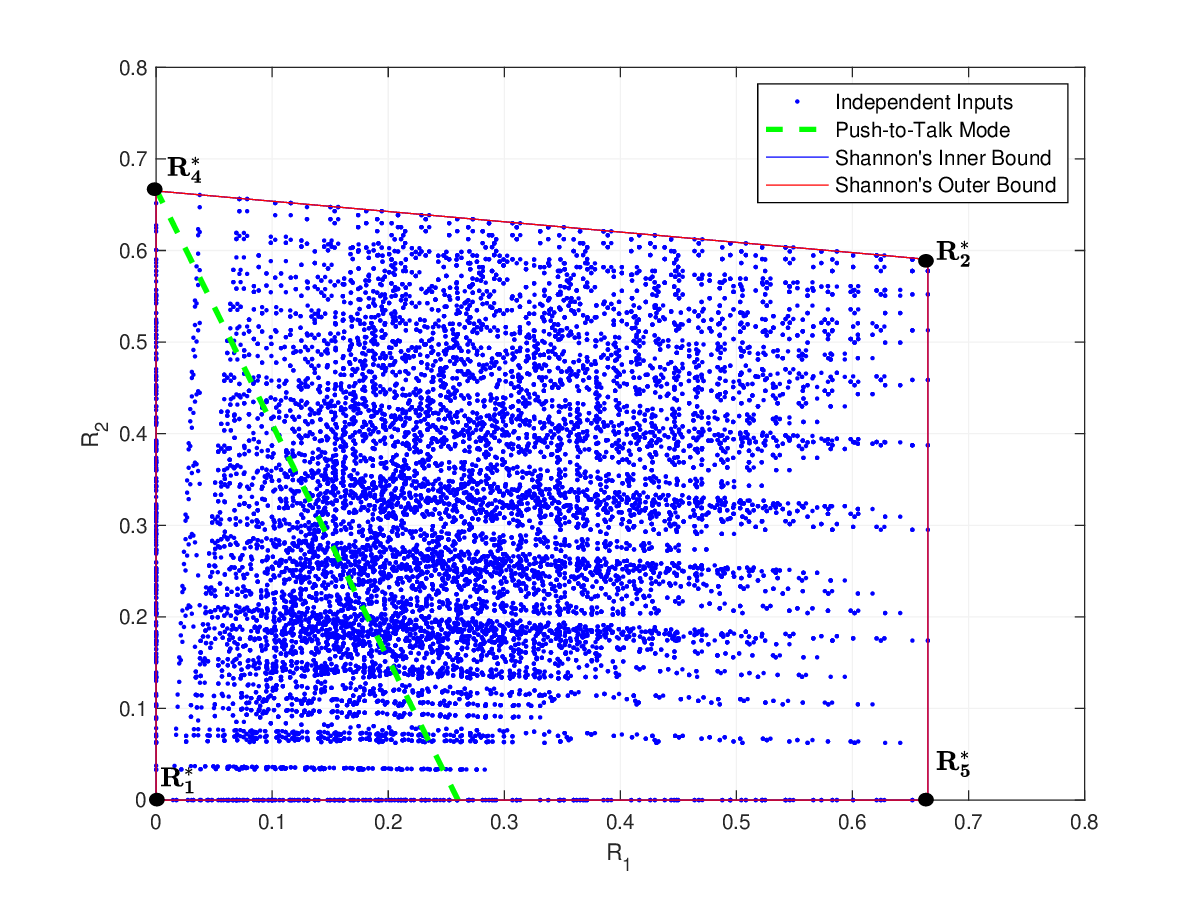}
\label{fig:3}}
\hfil
\subfigure[Setting 4 (unused rate pairs: $\bm{R}_3=(0.2601, 0)$ and $\bm{R}_4=(0, 0.0791)$)]{\includegraphics[draft=false, width=3in]{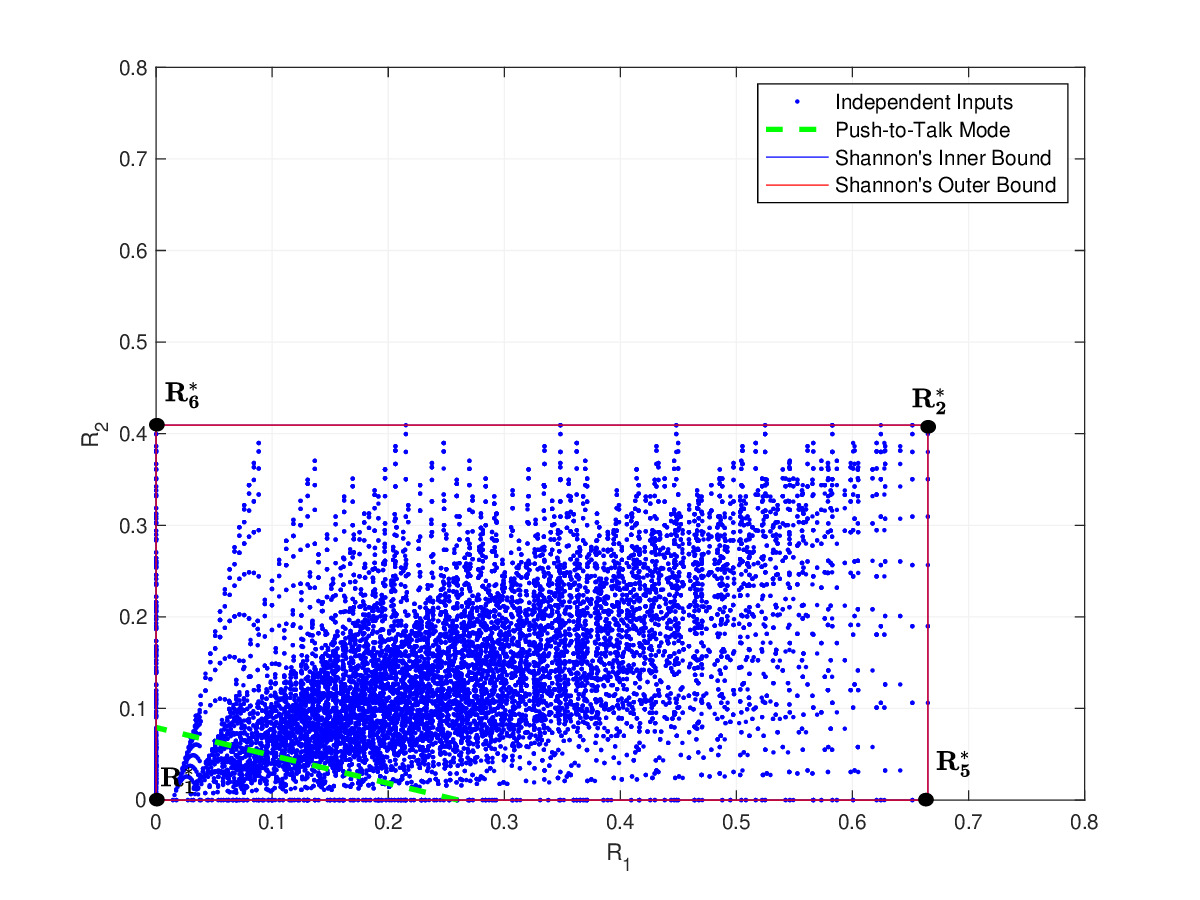}
\label{fig:4}}
}
\setlength{\belowcaptionskip}{-10pt}
\caption{The capacity region of the generalized DM-PTT-TWCs in Table~II. Except for Setting 1, the capacity region is determined by four rate pairs.}
\label{fig:main}
\end{figure*}

In Figures 2(a)-(d) (corresponding to Settings 1--4, respectively), the blue dots\footnote{In our computations, we discretized the standard 2-dimensional simplex to generate the input distributions for each user. The mutual information $I(X_j; Y_k|X_k)$ is then evaluated under the product of the discretized input distributions. A similar approach is used to obtain rate pairs in Shannon's outer bound region.} are the achievable rate pairs via independent inputs of the form: $P_{X_1, X_2}=P_{X_1}\cdot P_{X_2}$; Shannon's inner bound region $\mathcal{C}_\text{I}$ is then given by taking the convex hull of those rate pairs.  
Shannon's outer bound $\mathcal{C}_\text{O}$ is obtained using a similar method, but the convex hull operation is not needed. 
We also depict the achievable rate region using the half-duplex transmission mode (via input symbol `$0$'). 
In all settings, we have that $\mathcal{C}_\text{I}=\mathcal{C}_\text{O}$ as expected. 

In Figure~\ref{fig:1}, we first observe that the half-duplex transmission can attain the entire capacity region. 
Indeed, although full-duplex transmission is reliable, the large difference between $C_{j, 0}$ and $C_{j, x_k}$ (for $x_k\neq 0$) limits the rates achievable via two-way simultaneous transmission and hence the half-duplex transmission is still optimal (in the sense of achieving capacity).  
Nevertheless, the benefit of full-duplex transmission can be made significant by increasing the value of $C_{j, x_k}$ for $x_k\neq 0$. 
In Figure~\ref{fig:2}, we illustrate a situation where two-way simultaneous transmission achieves better rate pairs than using the half-duplex transmission. 

Moreover, when the $C_{j, x_k}$'s ($x_k\neq 0$) are much larger than $C_{j, 0}$, using $[P_{Y_2|X_1, X_2}(\cdot|\cdot, 0)]$ and $[P_{Y_1|X_1, X_2}(\cdot|0, \cdot)]$ for information transmission becomes inefficient since they contribute very little to the overall transmission rates in \eqref{eq5} and \eqref{eq6}. 
In this case, one should expect to abandon the (relatively) inefficient channels and use only the efficient ones. 
This is illustrated in Figs.~\ref{fig:3} and~\ref{fig:4}.
In an extreme case, such as Setting 4, the upper-right corner point of the capacity region is given by $\bm{R}^*_2=(0.6667, 0.4105)=(C_{1, 1}, C_{2, 1})$, implying that both users shut down the state-dependent one-way channels $[P_{Y_2|X_1, X_2}(\cdot|\cdot, 0)]$ and $[P_{Y_1|X_1, X_2}(\cdot|0, \cdot)]$ and only use the remaining channels for information exchange.

\section{Conclusions}
We identified a channel symmetry property under which Shannon's capacity inner bound is tight for a class of generalized DM-PTT-TWCs.
This symmetry property differs from prior ones in that it necessitates the use of a time-sharing scheme to achieve capacity.
Specifically, a time-sharing coding scheme that involves two independent transmissions is optimal. 
Viewing the generalized DM-PTT-TWC as two sets of one-way channels, we further observed that one-way channel components with (relatively) low capacity should be abandoned for efficient transmissions.  
Future research directions include finding a more general tightness condition for DM-TWCs, identifying the connections between different channel symmetry properties (in particular between the channel symmetry property introduced in this paper and the ones in \cite{chaaban2017} and \cite{jjw2018}), and investigating the transmission of correlated sources over the generalized DM-PTT-TWCs. 

\begin{appendix}
The appendix establishes input-output mutual information results for one-way channels that are of the same type as the state-dependent one-way channels in the generalized PTT-TWC of Theorem~1. 
Let $\mathcal{X}=\{0, 1, \dots, r-1\}$ and $\mathcal{Y}=\{0, 1, \dots, s-1\}$ denote channel input and output alphabets, respectively, for some integers $r\ge 3$ and $s\ge 2$.
Suppose that the set of probability vectors $\{[P_{Y|X}(\cdot|x_1)]: x_1\in\mathcal{X}\backslash \{0\}\}$ specifies a weakly-symmetric channel and $P_{Y|X}(y|0)=1/s$ for all $y\in\mathcal{Y}$. 
The input-output mutual information for a specific channel input symbol $x\in\mathcal{X}$ is defined as
\begin{IEEEeqnarray}{rCl}
I(X=x; Y)\triangleq \sum_{y\in\mathcal{Y}}P_{Y|X}(y|x)\cdot\log\frac{P_{Y|X}(y|x)}{P_Y(y)}.\nonumber
\end{IEEEeqnarray} 
The following results are needed in the proof of Theorem~1. 

\begin{lemma}\label{lma:capacity}
The capacity of the channel with the above properties is given by $C^*=\max_{P_X} I(X; Y)=\log s-H([P_{Y|X}(\cdot|1)])$, where $H([P_{Y|X}(\cdot|1)])$ denotes the entropy of the probability vector $[P_{Y|X}(\cdot|1)]$. 
The capacity-achieving input distribution is given by:  
\begin{equation}
P^*_{X}(x) =
\begin{cases}
0 & \text{if } x=0,
\\
\frac{1}{r-1} & \text{otherwise}.
\end{cases}\nonumber
\end{equation}
\end{lemma}
\begin{IEEEproof}
We apply the KKT condition for channel capacity \cite[Theorem~4.5.1]{gallager1968} to check the optimality of $P^*_{X}$. 
Under $P^*_X$, we first have that $I(X=x; Y)=\log s-H([P_{Y|X}(\cdot|1)])$ for $x\neq 0$ \cite[Theorem~7.2.1]{cover2012} since $P_X^*$ is a uniform distribution when restricted to the input alphabet $\mathcal{X}\setminus\{0\}$ and the channel with the restricted inputs is weakly-symmetric.   
Moreover, for $x\neq 0$, we have
\begin{IEEEeqnarray}{rCl}
I(X=0; Y)&=& \sum_{y=0}^{s-1} \frac{1}{s}\cdot\log\frac{1/s}{\sum_{x'\neq 0}P_{Y|X}(y|x')/(r-1)}\nonumber\\
&=& \log\frac{r-1}{s}-\sum_{y=0}^{s-1}\frac{1}{s}\cdot\log\left(\sum_{x'\neq 0}P_{Y|X}(y|x')\right)\nonumber\\
&=& \log\frac{r-1}{s}-\log\left(\sum_{x'\neq 0}P_{Y|X}(y'|x')\right)\label{appdix:1}\\
&=& -\log s +\log(r-1) -\underbrace{\left(\sum_{y'=0}^{s-1} P_{Y|X}(y'|x)\right)}_{=1}\cdot\log\left(\sum_{x'\neq 0}P_{Y|X}(y'|x')\right)\nonumber\IEEEeqnarraynumspace\\
&\le & -H(Y|X=x)+\log(r-1)-\sum_{y'=0}^{s-1} P_{Y|X}(y'|x)\cdot\log\left(\sum_{x'\neq 0}P_{Y|X}(y'|x')\right)\label{appdix:2}\IEEEeqnarraynumspace\\
&=& \sum_{y'=0}^{s-1} P_{Y|X}(y'|x)\cdot\log\frac{P_{Y|X}(y'|x)}{\sum_{x'\neq 0}P_{Y|X}(y'|x')/(r-1)}\nonumber\\
&=& I(X=x; Y),\nonumber
\end{IEEEeqnarray}
where $y'\in\mathcal{Y}$ is arbitrary in \eqref{appdix:1} since $\sum_{x'\neq 0}P_{Y|X}(y|x')$ does not depend on $y$ and \eqref{appdix:2} holds since $H(Y|X=0)\le \log s$. 
Combining the above results then gives that $I(X=0; Y)\le I(X=x; Y)$ for all $x\neq 0$, thus implying the optimality of $P^*_{X}$.
Finally, we conclude that $C^*=\max_{P_X} I(X; Y)=I(X=x; Y)$ for any $x\neq 0$ by the KKT condition. 
\end{IEEEproof}

\begin{lemma}\label{lma:rateloss}
For any $0\le \alpha \le 1$, consider the following channel input distribution:
\[
P^{(1)}_{X}(x) =
\begin{cases}
\alpha & \text{if } x=0,
\\
\frac{1-\alpha}{\mu-1} & \text{otherwise},
\end{cases}
\]
Let $P^{(2)}_{X}$ denote any input distribution with $P^{(2)}_{X}(0)=\alpha$. Then, we have that $I^{(2)}(X; Y)\le I^{(1)}(X; Y)=(1-\alpha)\cdot C^*$ (here the superscript indicates which input distribution is used for evaluation). 
\end{lemma}
\begin{IEEEproof}
First, we have that $H^{(2)}(Y)\le \log s=H^{(1)}(Y)$. 
Also, since $H(Y|X=x)=H(Y|X=1)$ for all $x\neq 0$ due to the weakly-symmetric structure, one can easily conclude that $H^{(1)}(Y|X)=H^{(2)}(Y|X)$.
The above results then imply that $I^{(2)}(X; Y)\le I^{(1)}(X; Y)$. 
Moreover, a direct computation (with the result in Lemma~2) yields that $I^{(1)}(X; Y)=(1-\alpha)\cdot C^*$, thereby completing the proof. 
\end{IEEEproof}
\end{appendix}

\bibliographystyle{IEEEtran}
\bibliography{thesis}

\end{document}